\documentclass[aps,prl,twocolumn,groupedaddress,showpacks]{revtex4-2}

\usepackage{graphicx}
\usepackage{dcolumn}
\usepackage{bm}
\usepackage[colorlinks=true,urlcolor=blue,linkcolor=blue,citecolor=blue]{hyperref}

\begin{document}


\title{Supercondutivity of Nb-Ta-Ti-Zr-Hf high entropy alloy polycrystalline and amorphous thin films}




\author{P. Hru\v ska}
\affiliation{Institute of Physics of the CAS, Na Slovance 2, 182 21 Prague 8, Czech Republic}

\author{Z. Jan\accent23 u}
\affiliation{Institute of Physics of the CAS, Na Slovance 2, 182 21 Prague 8, Czech Republic}


\author{J. \v C\' i\v zek}
\affiliation{Faculty of Mathematics and Physics, V Holesovickach 2, 180 00 Prague 8, Czech Republic}

\author{F. Luk\' a\v c}
\affiliation{Institute of Plasma Physics of the CAS, Na Slovance 2, 182 21 Prague 8, Czech  Republic}

\date{\today}

\begin{abstract}
We studied the superconductivity of high-entropy Hf-Nb-Ta-Ti-Zr alloy films affected by structural order. While films deposited from the same target on a substrate with a room temperature are amorphous, and if superconducting, only at temperatures below 1.9 K, films deposited on a substrate with a temperature of 740 $^o$C and 630 $^o$C are superconducting with a critical temperature of 6.61 K and 6.63 K, respectively. The observed effect of structural order on superconductivity can be explained by the theory of superconductivity in strongly bound amorphous materials recently published by Baggioli et al.
\end{abstract}


\maketitle

\section{Introduction}

High entropy alloys (HEA) are a new class of materials with remarkable mechanical and electronic properties. In Hf-Nb-Ta-Ti-Zr alloys, all elements are superconducting. All bulk samples of these alloys prepared so far have a critical temperature lower than pure Nb, the commercially used Nb-Ni alloy, and the compound Nb$_3$Sn \cite{Wilson2008}. HEA superconductors are type II. Parameters such as critical temperature, critical fields, or critical current density do not exhibit a cocktail effect \cite{Yeh2006}. A maximum $T_c = 8.03$ K was observed in bulk (TaNb)$_{0.7}$(ZrHfTi)$_{0.3}$ with an upper critical field $B_{c2}=6.67$ T, while the maximum $B_{c2}=11.67$ T was observed in (TaNb)$_{0.5}$(ZrHfTi)$_{0.5}$ with $T_c = 6.46$ K \cite{Rohr2016}. Critical temperature $T_c\approx 7.3$ K, $B_{c2}\approx 8.15$ T at 0 K with a temperature dependence $B_{c2}(T)/B_{c2}(0)=(1-(T/T_c)^{1.51})$ and the lower critical field $B_{c1} \approx 32$ mT at 2 K is reported for bulk Ta$_{34}$Nb$_{33}$Hf$_{8}$Zr$_{14}$Ti$_{11}$ in Ref. \cite{Kozelj2014}. A study of HEA films with the composition (TaNb)$_{1-x}$(ZrHfTi)$_x$ and a thickness of 600-950 nm, prepared by magnetron sputtering, is presented in Ref. \cite{Zhang2020}. The highest $T_c = 6.9$ K was obtained for $x \approx 0.43$ and the highest upper critical field $B_{c2} = 11.05$ T for $x \approx 0.65$. Another study of Ta–Nb–Hf–Zr–Ti HEA films produced by pulsed laser deposition on a $c-$cut Al$_2$O$_3$ substrate reports the highest $T_c=7.28$ K and critical current density $j_c > 10$ GA m$^{-2} $ at 4.2 K \cite{Jung2022}.

We prepared HEA films of the same chemical composition, from the same target, and found that their parameters strongly depend on the temperature of the substrate on which the film is deposited. Diffraction analysis shows that the films deposited on substrates at room temperature are amorphous or with little structural order. If these films are superconducting, then their critical temperature is less than 1.9 K. On the other hand, films deposited on substrates with a temperature of 740 $^o$C and 630$^o$C have structural order and are superconducting with $T_c$ of 6.61 K and 6.63 K, respectively.

It is known that good crystalline superconductors are insensitive to dilute non-magnetic impurities, but a significant disorder effect becomes apparent when the lattice structure becomes amorphous \cite{Anderson1959,Baggioli2020}. With few exceptions, the critical temperature $T_c$ decreases with increasing disorder.

\section{Film deposition and characterisation}

Four Hf-Nb-Ta-Ti-Zr films were deposited on MgO (100) substrates by DC magnetron sputtering with a single target of 20 mm diameter. The deposition conditions are detailed in the table \ref{TableDeposition}. The depositions took place in a UHV chamber with a base pressure of $10^{-5}$ Pa and a lower, subsequently filled with Ar at a deposition pressure of 2 Pa. The magnetron power was set to 10 W. Film thicknesses were estimated based on a deposition rate of 5 nm/min, resulting in $d=350(30)$ nm. In addition, an approximately 30 nm thick aluminum cover layer was deposited on each Hf-Nb-Ta-Ti-Zr film to protect it against oxidation in the ambient atmosphere.

\begin{table}
\caption{\label{TableDeposition} Deposition conditions}
\begin{ruledtabular}
\begin{tabular}{ccccc}
sample & $p_0$ (Pa) & target & $T_{d}$ ($^o$C) & $P_{heat}$ (W)\\
\hline
HSC1 & $1.3 \times 10^{-5}$ & HEA1 & 20 & 0 \\
HSC2 & $2.1 \times 10^{-5}$ & HEA1 & 740 & 100 \\
HSC3 & $1.2 \times 10^{-7}$ & HEA2 & 20 & 0 \\
HSC4 & $4.6 \times 10^{-8}$ & HEA2 & 630 & 78 \\
\end{tabular}
\end{ruledtabular}
\end{table}

After the HSC2 deposition, the HEA1 target was consumed, as evidenced by the visible hole in the target. Consequently, the film was inadvertently contaminated by indium, utilized as an adhesive to affix the target to a copper target holder. Subsequently, for the depositions of HSC3 and HSC4 films, a fresh HEA2 target was employed. Additionally, a chamber bakeout enhanced the base pressure by two orders of magnitude.

Depositions of HSC1 and HSC3 films were carried out at room temperature. For the depositions of HSC2 and HSC4 films, the MgO substrate underwent heating, reaching temperatures $T_d$ of 740 $^o$C and 630 $^o$C respectively. It is important to note that, during the deposition of the HSC4 film, the temperature could not be increased to the same value as for the HSC2 film due to shortcuts in the heating apparatus at higher heating currents. Deposition at room temperature typically yields films with an amorphous structure, whereas deposition at elevated temperatures results in a nanocrystalline structure.

\subsection{Morphology of films}

\begin{figure}
\center
\includegraphics[scale=1.2,angle=90]{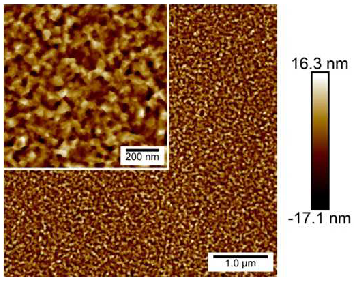}
\caption{\label{HSC1hs2001000}
Morphology of the HSC1 sample deposited at $T_d$ = 20 $^o$C. (height sensor 0.2 and 1 $\mu$m): rms = 4.9 nm/4.8 nm.}
\end{figure}

\begin{figure}
\center
\includegraphics[scale=1.2,angle=90]{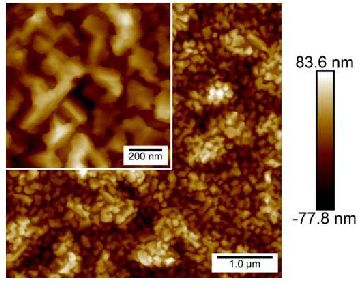}
\caption{\label{HSC2hs2001000}
Morphology of the HSC2 sample deposited at $T_d$ = 740 $^o$C. (height sensor 0.2 and 1 $\mu$m): rms = 4.9 nm/4.8 nm.}
\end{figure}

\begin{figure}
\center
\includegraphics[scale=1.2,angle=90]{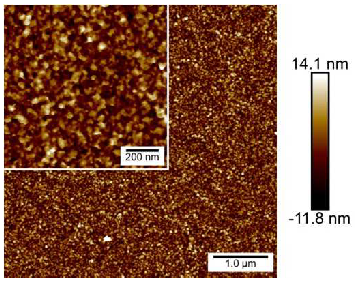}
\caption{\label{HSC3hs2001000}
Morphology of the HSC3 sample deposited at $T_d$ = 20 $^o$C. (height sensor 0.2 and 1 $\mu$m): rms = 4.9 nm/4.8 nm.}
\end{figure}

\begin{figure}
\center
\includegraphics[scale=1.2,angle=90]{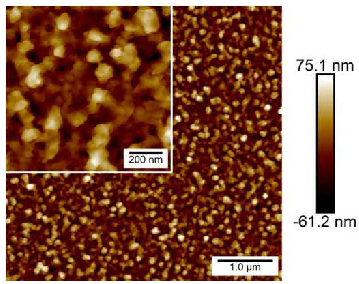}
\caption{\label{HSC4hs2001000}
Morphology of the HSC4 sample deposited at $T_d$ = 630 $^o$C. (height sensor 0.2 and 1 $\mu$m): rms (roughness) = 4.9 nm/4.8 nm.}
\end{figure}

The morphology of the films was characterized using AFM performed at room temperature in Peak Force Tapping mode, as shown in Figs. \ref{HSC1hs2001000},\ref{HSC2hs2001000},\ref{HSC3hs2001000}, and \ref{HSC4hs2001000}. Measurements included $1 \times 1$ $\mu$m$^2$ and $5 \times 5$ $\mu$m$^2$ scans with a resolution of $512 \times 512$ points for each film. Root mean square roughnesses were subsequently calculated.

The morphology obtained by AFM features Al nanocrystals growing on distinct films. The smooth surface of the films deposited at room temperature results from Al growth on the amorphous film. In contrast, the surface roughness of films deposited at elevated temperatures is 3 to 5 times higher, attributable to the nanocrystalline nature of the film. Example AFM scans of reference
films without Al cover are shown below.

\subsection{Structure of films}

\begin{figure}
\center
\includegraphics[scale=1.2,angle=90]{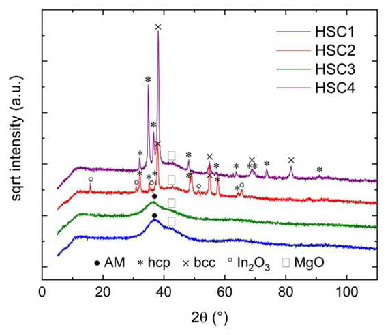}
\caption{\label{XRD HSC fig2b}
XRD spectra, shifted vertically for clarity, show crystalline structure of samples HSC2 and HSC4 and amorphous structure of samples HSC1 and HSC3.}
\end{figure}

XRD measurements were carried out in Bragg-Brentano $\theta-2\theta$ geometry. The XRD patterns of the films are presented below. An asymmetric scan geometry was employed to eliminate the MgO (100) reflection, with its residuum indicated by an open square. XRD analysis confirmed the amorphous structure of HSC1 and HSC3 films deposited at room temperature, identified by a broad peak at 36.8$^o$ marked by a full circle, as shown in Fig. \ref{XRD HSC fig2b}. Films deposited at elevated temperatures exhibited a nanocrystalline structure characterized by a mixture of two intermetallic phases: hcp phase rich in Zr and Hf, and bcc phase rich in Nb and Ta. Indium contamination of the HSC2 film was evident from the presence of additional peaks, preliminarily identified as In$_2$O$_3$. Detailed XRD spectra of HSC2 and HSC4 films are provided for comparison.

\section{Superconductivity of films}

Samples with the square cross-section of the side $a=4$ mm and film thickness of $d=350$ nm were measured non-contact using a SQUID magnetometer with continuous reading \cite{Janu2017}. The critical temperature was determined from a measurement of the temperature dependence of the ac magnetic moment. Measurements showed that nanocrystalline samples HSC2 and HSC4 are superconducting with critical temperatures $T_c$ of 6.61 K and 6.63 K, respectively, while amorphous samples HSC1 and HSC3, if they are superconducting at all, have a critical temperature lower than 1.9 K, as determined from measurements on the QD MPMS.

\subsection{Critical depinning current density}

The samples HSC2 and HSC4 are type II superconductors. Their critical depinning current density and its temperature dependence were determined from the measured temperature dependence of the ac magnetic moment, namely from the real part of the Fourier coefficient at the fundamental frequency, using a model for the ac susceptibility of thin disks in the critical state placed in a transverse magnetic field \cite{Brandt1993,Clem1993,Clem1994,Brandt1997,Brandt1998,Brandt1998B}. The analytical model for disks can also be used well for samples with a square cross-section, because the result only differs by 0.1 \% \cite{Brandt1997,Brandt1998}. The model is valid if $d \ge \lambda$, or, if not, then if $\lambda^2 \ll Rd/2$, where $\lambda$ is the London flux penetration length and $R$ is the disk radius. Due to the stray field of currents induced in a disk in a transverse field, the lower critical field $B_{c1}$ at the edge of the disk is exceeded already in the applied field $B_a>(d/R)^{1/2}B_{c1}$ \cite{Clem1994,Chen2010}. For $d=350$ nm, $R=2$ mm and $B_{c1}=1$ mT, this field is $(d/R)^{1/2} B_{c1} \approx 13$ $\mu$T, which is of the order of magnitude of the Earth's field.

\begin{figure}
\center
\includegraphics[scale=0.55,angle=0]{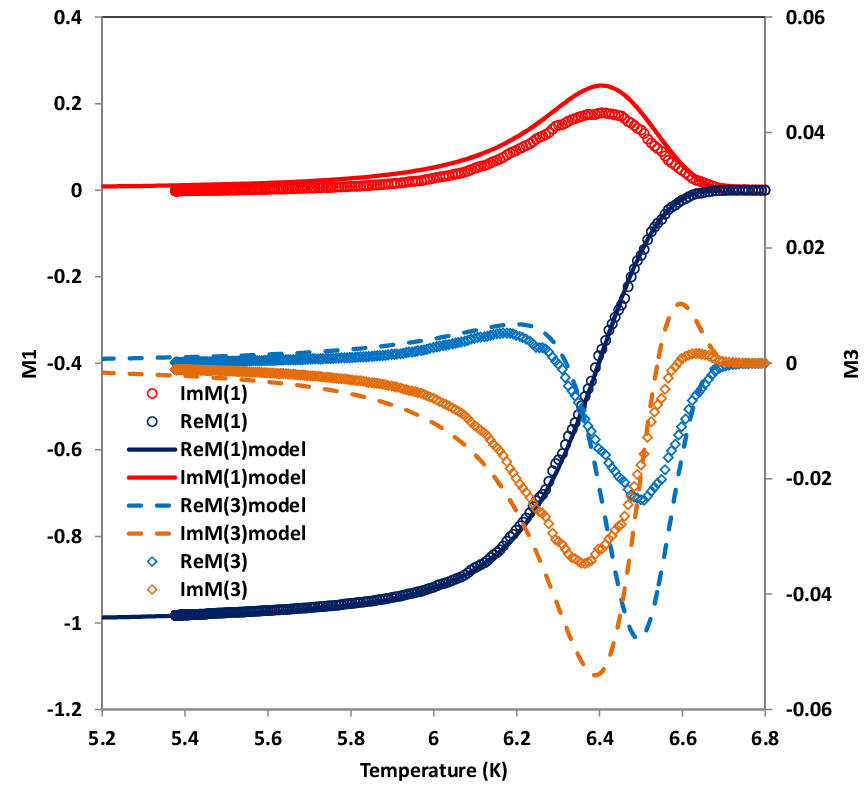}
\caption{\label{M(T)}
Normalized experimentally determined magnetic moment coefficients (symbols) at the fundamental frequency and the third harmonic frequency of the HSC4 sample, measured in an ac field with amplitude $B_p = 0.1$ mT and frequency 1.5625 Hz, and model-calculated coefficients (lines) for the disks. Experimental data were measured at warming rate of 1 K/min.
}
\end{figure}

\begin{figure}
\center
\includegraphics[scale=0.55,angle=0]{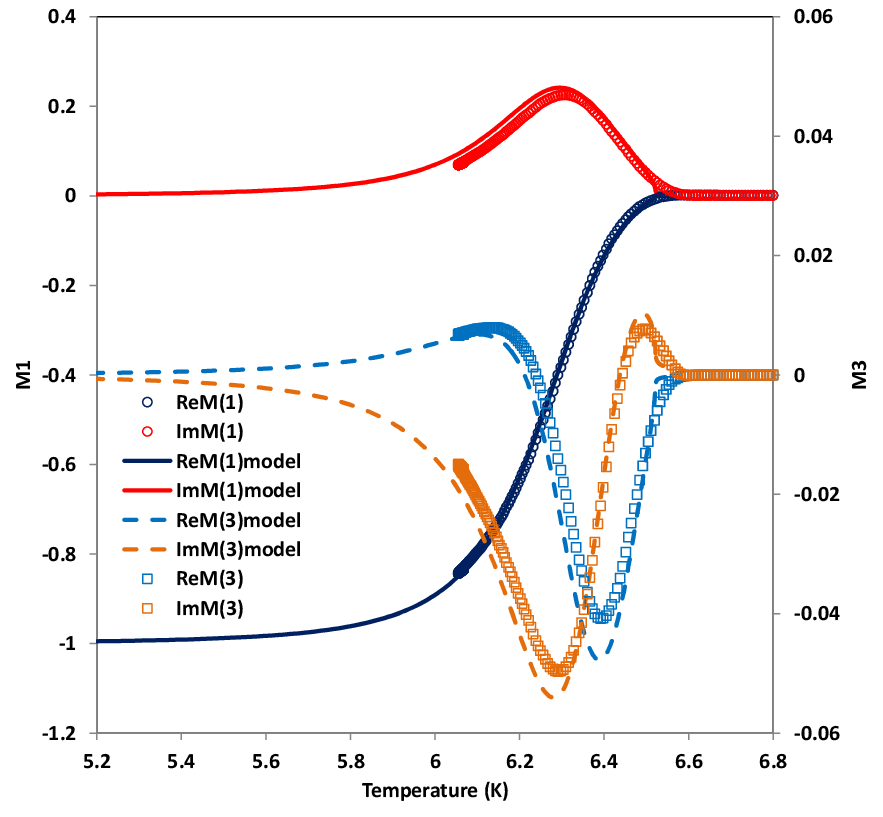}
\caption{\label{M(T)2}
Normalized experimentally determined magnetic moment coefficients (symbols) at the fundamental frequency and the third harmonic frequency of the HSC2 sample, measured in an ac field with amplitude $B_p = 0.1$ mT and frequency 1.5625 Hz, and model-calculated coefficients (lines) for the disks. Experimental data were measured at warming rate of 1 K/min.
}
\end{figure}

Based on the model, we calculated the magnetization loops and their Fourier coefficients, a set of data pairs $[L_i,\mathcal{M}_{ki}]_m$, where $L=\ln(B_p/B_f^*)$, $B_p$ is the amplitude of the applied ac field, $B_f^*=\mu_0 J_c d/2$ is the characteristic field analogous to a full-penetration field $B_f=\mu_0 J_c R$ of long cylinders, $\mathcal{M}_k$ are the Fourier coefficients of the ac magnetic moment, and $k$ denotes the $k$-th harmonic frequency \cite{Clem1994,Janu2017}. The coefficients are normalized so that for $B_p/B_f^* \rightarrow 0$ is the real part of the coefficient at the fundamental frequency, the moment of supercurrent being in phase with the applied field, $\mathcal{M}_1'=-1$. For the set of the data pairs $[\mathcal{M}_{1i}',L_i]_m$ 
we found the approximating function

\begin{equation}\label{SigmoidDiskB(ReM1)}
L(\mathcal{M}_1')=C-D\ln\left(B/\left(\mathcal{M}_1'+A\right)-1\right),
\end{equation}

\noindent where the parameters are $A=1.0026(4)$, $B=0.9976(6)$, $C=0.444(1)$, and $C=0.538(1)$, standard error $\sigma=0.004$ \cite{Janu2025}. Using this function, we solve the inverse problem, converting the measured temperature dependence of the magnetic moment, a set of data pairs $[T_i,\mathcal{M}_{1i}']_e$, to the temperature dependence of the characteristic field, a set of data points $[T_i,L_i]_e$, where $L_i=L(\mathcal{M}_{1i}')$. Note that the experimentally obtained coefficients are also normalized. Then for temperature $T_i$ the characteristic field is $B_{fi}^* = B_p \exp(-L_i)$ and the critical depinning current density is $J_{ci} = 2 B_{fi}^* / \mu_0 d$. To assess the validity of the model used to solve the inverse problem, we plot the temperature dependence of the experimental and modeled complex coefficients $\mathcal{M}_1$ and $\mathcal{M}_3$.  Fig. \ref{M(T)} shows the ac moment coefficients of the sample HSC4 measured in a sinusoidal field with amplitude $B_p=0.1$ mT and frequency $f=1.5625$ Hz. The same is shown in Fig. \ref{M(T)2} for sample HSC2. The correlation between the experimental and theoretical data is evident, including the 3rd harmonic, which is characteristic of the critical state and is absent in the flux-flow state. The effective temperature $T_e$ for the model data is obtained by finding some approximation function $T(L)$ for the set of experimentally obtained data points $[L_i,T_i]_e$, the inverse function of the temperature dependence of the characteristic field, and calculating the temperatures $T_{ei}=T(L_i)$ for the values of $L_i$ for which the magnetization curves were calculated.
The determined parameters are listed in table \ref{Table0}.

The model predicts $\max \mathcal{M}_1''=0.241$ which occurs at $B_p = 1.942 B_f^* = 0.971 \mu_0 J_c d$ \cite{Clem1994}. For the sample HSC4, the maximum $\mathcal{M}_1''=0.183$ in the experimental data set occurs at the temperature $T_{f}=6.41$ K. For the sample HSC2, the maximum $\mathcal{M}_1''=0.226$ in the experimental data set occurs at the temperature $T_{f}=6.30$ K.

\begin{table}
\caption{\label{Table0}Parameters determined for an applied field with amplitude $B_p=0.1$ mT and frequency $f=1.5625$ Hz. 
Please note that the temperature sensor is not calibrated to mK accuracy, but the resolution is better than mK \cite{Janu2017}.}
\begin{ruledtabular}
\begin{tabular}{lllll}
sample   &               & HSC2 & HSC4 & Nb \\
\hline
$T_c$    & [K]           & 6.610(7)     & 6.63(1) & 8.897(1) \\
$B_f(0)$ & [mT]          & 3.9          & 1.1     & 1.57 \\
$J_c(0)$ & [GA m$^{-2}$] & 18(7)        & 5.2(3)  & 10.1(9) \\
$m$      &               & 2.8(7)       & 4.3(4)  & 5.8(6) \\
$n$      &               & 2.12(8)      & 1.55(8) & 1.22(2) \\
\end{tabular}
\end{ruledtabular}
\end{table}

\begin{figure}
\center
\includegraphics[scale=0.55,angle=0]{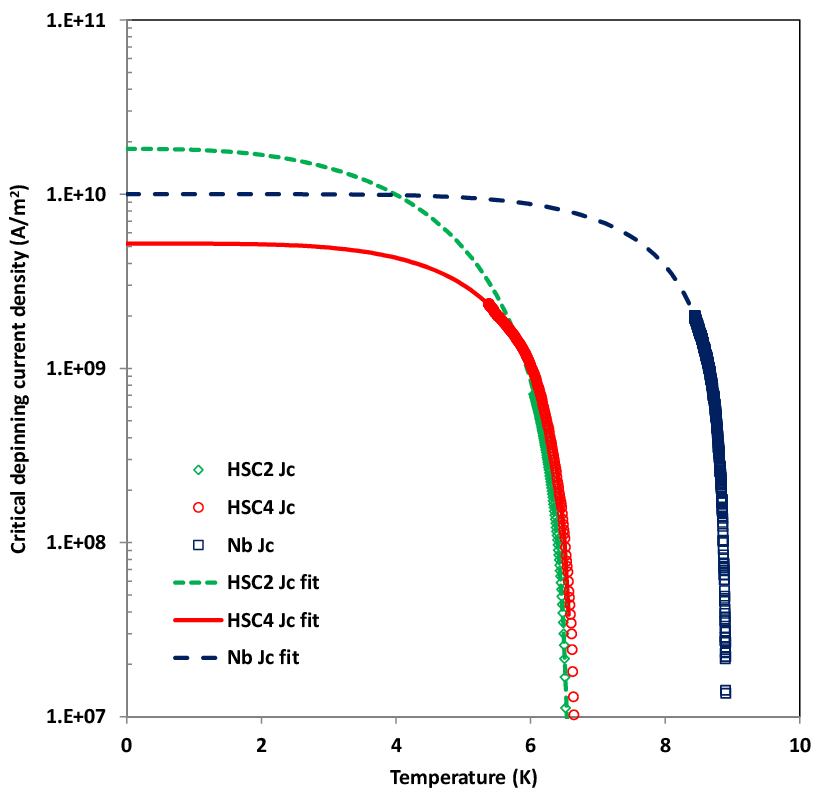}
\caption{\label{jc(T)}
Critical depinning current densities of samples HSC2 and HSC4 (symbols) as a function of temperature, as determined from AC field measurements. For comparison, the critical depinning current density of the Nb film, which was determined by the same method, is also shown. The lines are for plotting approximation functions.
}
\end{figure}

The commonly used form of the temperature dependence of the critical depinning current density is

\begin{equation}\label{CriticalCurrentDensityVsT}
J_c(T)/J_c(0) = \left(1-\left(T/T_c\right)^m\right)^n,
\end{equation}

\noindent where 
the exponents are $m \approx 2$ and $1 < n < 3$, as is usually observed in experiments \cite{Clem1993,Campbell1972}. 
Fitting Eq. \ref{CriticalCurrentDensityVsT} to the set of data $[T_i,J_{ci}]$ we obtain extrapolated $J_c(0)$ and exponents. The temperature dependence of the critical current density is shown in Fig. \ref{jc(T)} including, for comparison, the temperature dependence of the critical current density in a 250 nm thick Nb film 
\cite{Janu2015}.

\begin{figure}
\center
\includegraphics[scale=0.55,angle=0]{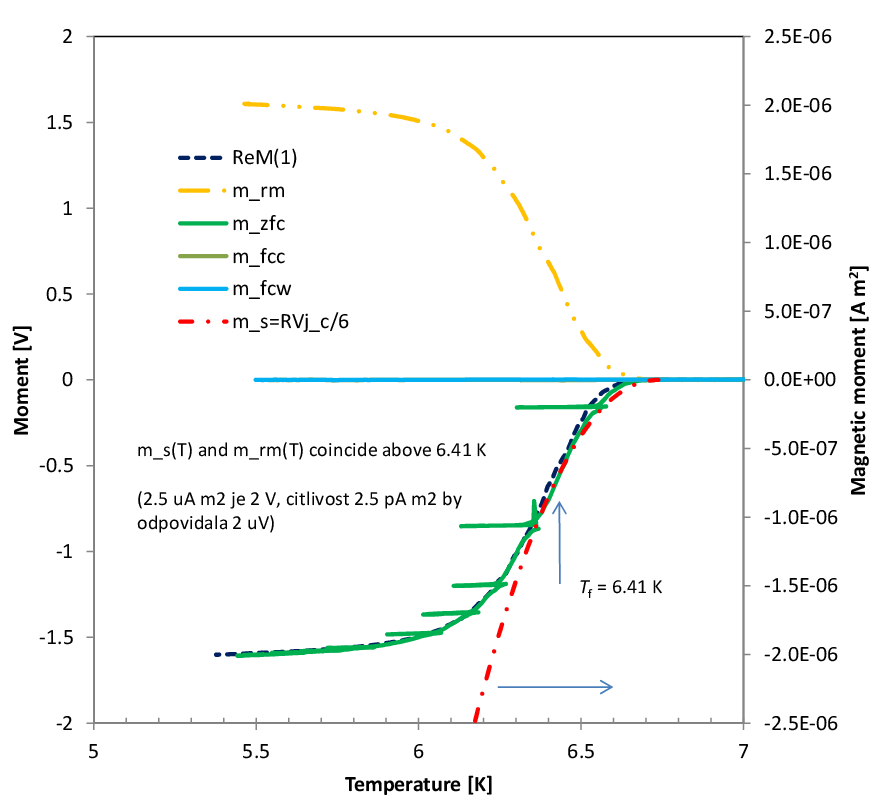}
\caption{\label{m_dc(T)}
Temperature dependence of the dc moment (FCC, FCW, ZFC and RM procedures), the coefficient $\mathcal{M}'_1$, and the saturated moment $m_s$ calculated from $J_c(T)$ for the HSC4 sample. For the temperature $T > T_f= 6.41$ K, the ZFC and RM moments are saturated. The horizontal lines in the temperature dependence of $m_{zfc}$ prove the critical state.
}
\end{figure}

Fig. \ref{m_dc(T)} shows the temperature dependence of the dc magnetic moment. The  magnetic moments $m_{fcc}(T)$ and $m_{fcw}(T)$ measured during cooling (FCC) and warming (FCW) in the applied dc field $B_{dc} =0.1$ mT do not change (unmeasurable) during the transition from the normal state to the superconducting state. This indicates that effective vortex pinning just below $T_c$ prevents the expulsion of vortices that spontaneously form in the superconducting condensate during the transition from the normal to the superconducting state \cite{Stephens2002,Kibble2003}. (But $B_a > (d/R)^{1/2} B_{c1}$) The ZFC moment $m_{zfc}$ of induced persistent shielding currents was recorded while warming the sample after cooling it in zero field to a temperature of 5.4 K and switching on the field $B_{dc} = 0.1$ mT. The remanent moment $m_{rm}$ of induced persistent shielding currents was recorded during warming after FCC to a temperature of 5.4 K and switching of the field. The magnitudes of these moments are almost the same, since $m_{rm}(T) + m_{zfc}(T ) = m_{fcc}(T)$ and $m_{fcc}(T) \ll m_{zfc}(T)$ \cite{Clem1993}. The saturated moment $m_s(T) = - J_c(T)(a/2)V/3$ is calculated for $J_c(T)$ determined from the ac measurement. The $m_{zfc}(T)$ and $m_s(T)$ curves merge above the temperature $T_f=6.41$ K, which is the temperature at which ac loss $\mathcal{M}_1''B_p/2$ has a maximum. Note that the dc field and the ac field amplitude are the same. Above $T_f$ the film is fully penetrated and the moments depend only on $J_c(T)$, not on the current density distribution.

During the measurement of the $m_{zfc}(T)$ moment, the temperature was partially lowered several times and then the warming continued again. The measurement shows that the moment (i.e., the current density distribution) remembers the highest temperature to which the sample was exposed and does not change when the temperature is lowered. This temperature hysteresis again proves the critical state.

The dc moment $m_{zfc}(T)$ and the ac moment $\mathcal{M}'_1(T)$ induced by the same change in the applied field, $B_p=B_{dc}$, do not depend on the rate of change of the field, that again indicates the quasi-static behavior characteristic of the critical state.


\section{Discussion}

All the films studied here have the same chemical composition, but differ in their structural order. The crystalline ones are superconducting with critical temperature of about 6.6 K, while the amorphous ones, if they are superconducting, have a critical temperature of less than 1.9 K. This result is consistent with the theory recently presented by Baggioli et al. \cite{Baggioli2020}. The HEA samples are conventional BCS superconductors. In structurally ordered materials, only longitudinal phonons contribute to the electron-phonon interaction. In amorphous materials, momentum is not conserved in electron-phonon scattering and transverse phonons also contribute to the pairing. Amorphous materials are strong coupling superconductors with an effective electron-phonon interaction described by the Eliashberg theory with the phonon spectral function dependent on the velocities of transverse and longitudinal phonons (sound velocity) and diffusion lengths (sound absorption coefficients) \cite{Baggioli2020,Marsiglio2020}. The Eliashberg spectral function of the phonons features two distinct peaks, which correspond to the low-frequency transverse and high-frequency longitudinal vibrational excitations. With increasing structural disorder, both peaks broaden, which corresponds to a decrease in ballistic phonon transport and an increase in the diffusivity of both longitudinal and transverse vibrational excitations, with the transverse mode in particular being sensitive to structural disorder. As the peak in the density of vibrational states becomes broader and shallower, the lifetime of the vibrational excitations becomes too short to allow for effective pairing/coupling.

Measuring the magnetic moment induced in an ac field as a function of temperature, commonly but incorrectly in this case called ac susceptibility measurement, allows for rapid sample characterization. One of the key parameters of type II superconductors is the critical depinning current density, which is usually determined from the vertical width of the magnetization loop measured at fixed temperature. In this case, it is necessary to know the absolute value of the magnetic moment. However, this can be a problem in the case of thin films. Another method is to determine the full penetration or characteristic field from which we obtain the critical depinning current density. The justification for using the model for measurement analysis is assessed as follows: a) the measured magnetic moment is independent of the rate of change of the applied field; b) non-zero coefficients for odd harmonics reflect the hysteresis loops \cite{Clem1994}.

The dependence used introduces the problem of uncertainty in the determination of $J_c(0)$ and exponent $m$ if the amplitude of the ac field is small. Then the relevant data are from a narrow interval below $T_c$ and these two parameters become intertwined. This is associated with uncertainty in their determination. Unfortunately, in virtually all experimental devices there is a limitation on the amplitude of the ac field. In addition, a small rate of change of the field is required so that the electric field induced in the sample does not cause flux flow.

Fig. \ref{M(T)} shows that in absolute value the experimentally determined coefficients $\mathcal{M}''_1(T)$, $\mathcal{M}'_3(T)$ and $\mathcal {M }''_3(T)$ are smaller than predicted by the model for thin disks in the critical state (true for HSC4 but not for HSC2). In a small applied ac field, the vortices pinned at the centers oscillate elastically without unpinning, so the losses are smaller than in the critical state, as are the coefficients for odd harmonics. On the other hand, a rapidly changing ac field induces an electric field that can cause the vortices to creep. Then the losses are larger than in the critical state and depend on the rate of change of the field, similar to the case of eddy currents \cite{Wilson2008,Campbell1972}.

\section{Conclusion}

We prepared Hf-Nb-Ta-Ti-Zr high-entropy alloy films with the same composition but different structural order. While films deposited from the same target on a room-temperature substrate are amorphous and, if superconducting, only at temperatures below 1.9 K, films deposited on a substrate with a temperature of 740 $^o$C and 630 $^o$C are superconducting with critical temperatures of 6.61 K and 6.63 K, respectively. This effect of structural order on superconductivity can be explained by the theory of superconductivity in strongly coupled amorphous materials. In this case, with increasing structural disorder, the localization of short-wavelength longitudinal phonons (important for superconductivity) increases and their lifetime decreases, leading to a decrease in the attractive retarded interaction between electron pairs. For superconducting films, we determined the critical depinning current density and its temperature dependence.

\section*{Acknowledgments}

Magnetic measurements at temperature below 4.2 K were carried out at the Low-Temperature Laboratory () using QD MPMS.

{}

\end{document}